\documentclass[twocolumn,prr,showpacs,amsmath,amssymb,superscriptaddress,nofootinbib,floatfix]{revtex4-2}

\usepackage{graphicx}
\usepackage{dcolumn}
\usepackage{bm}
\usepackage{epsfig}
\usepackage{natbib}
\usepackage{color}
\usepackage[normalem]{ulem}
\usepackage{enumerate}

\begin{document}

\title{The magnetic anisotropy of individually addressed spin states}

\author{L.\,C.\,J.\,M.\,Peters}
\affiliation{High Field Magnet Laboratory (HFML - EMFL), Radboud University, Toernooiveld 7, 6525 ED Nijmegen, The Netherlands}
\affiliation{Institute for Molecules and Materials, Radboud University, Heyendaalseweg 135, 6525 AJ Nijmegen, The Netherlands}

\author{P.\,C.\,M.\,Christianen}
\affiliation{High Field Magnet Laboratory (HFML - EMFL), Radboud University, Toernooiveld 7, 6525 ED Nijmegen, The Netherlands}
\affiliation{Institute for Molecules and Materials, Radboud University, Heyendaalseweg 135, 6525 AJ Nijmegen, The Netherlands}

\author{H.\,Engelkamp}
\affiliation{High Field Magnet Laboratory (HFML - EMFL), Radboud University, Toernooiveld 7, 6525 ED Nijmegen, The Netherlands}
\affiliation{Institute for Molecules and Materials, Radboud University, Heyendaalseweg 135, 6525 AJ Nijmegen, The Netherlands}

\author{G.\,C.\,Groenenboom}
\affiliation{Institute for Molecules and Materials, Radboud University, Heyendaalseweg 135, 6525 AJ Nijmegen, The Netherlands}

\author{J.\,C.\,Maan}
\affiliation{High Field Magnet Laboratory (HFML - EMFL), Radboud University, Toernooiveld 7, 6525 ED Nijmegen, The Netherlands}
\affiliation{Institute for Molecules and Materials, Radboud University, Heyendaalseweg 135, 6525 AJ Nijmegen, The Netherlands}

\author{E.\,Kampert}
\altaffiliation[Present address: ]{WMG, University of Warwick, Coventry CV4 7AL, U.K.}
\affiliation{High Field Magnet Laboratory (HFML - EMFL), Radboud University, Toernooiveld 7, 6525 ED Nijmegen, The Netherlands}
\affiliation{Institute for Molecules and Materials, Radboud University, Heyendaalseweg 135, 6525 AJ Nijmegen, The Netherlands}

\author{P.\,T.\,Tinnemans}
\affiliation{Institute for Molecules and Materials, Radboud University, Heyendaalseweg 135, 6525 AJ Nijmegen, The Netherlands}

\author{A.\,E.\,Rowan}
\altaffiliation[Present address: ]{Australian Institute for Bioengineering and Nanotechnology (AIBN), The University of Queensland, Brisbane, QLD 4072, Australia}
\affiliation{Institute for Molecules and Materials, Radboud University, Heyendaalseweg 135, 6525 AJ Nijmegen, The Netherlands}

\author{U.\,Zeitler}
\email[email: ]{Uli.Zeitler@ru.nl}
\affiliation{High Field Magnet Laboratory (HFML - EMFL), Radboud University, Toernooiveld 7, 6525 ED Nijmegen, The Netherlands}
\affiliation{Institute for Molecules and Materials, Radboud University, Heyendaalseweg 135, 6525 AJ Nijmegen, The Netherlands}


\begin{abstract}
Controlling magnetic anisotropy is a key requirement for the fundamental understanding of  molecular magnetism and is a prerequisite for numerous applications in magnetic storage, spintronics, and all-spin logic devices. In order to address the question of molecular magnetic anisotropy experimentally, we have synthesized single-crystals of a molecular spin system containing four antiferromagnetically coupled $s=5/2$ manganese(II) ions.  Using low-temperature cantilever magnetometry, we  demonstrate the selective population of the $S=0, 1, \ldots , 10$ spin states upon application of magnetic fields up to 33 T and map the magnetic anisotropy of each of these states. We observe a strong dependence of the shape and size of the magnetic anisotropy on the populated spin states, and, in particular, reveal an anisotropy reversal upon going from the lowest to the highest spin-state. 
\end{abstract}

\maketitle


The energy of a magnetic system with a given permanent or induced magnetic moment generally depends on the direction of an applied magnetic field. This so-called magnetic anisotropy results in a direction-dependent energy landscape with minima in certain directions (easy axes) and maxima in others (hard axes). The magnetization will preferentially orient itself along an easy axis and switching the magnetic moment requires crossing an energy barrier when passing through a hard axis. 
Magnetic anisotropy forms the basis for magnetic storage and the development of magnetic recording has triggered a strong effort to control its shape and size  \cite{Phys.Rev.B.054401},  a desire which has also extended to the applied fields of spintronics and all-spin logic devices \cite{Nat.Mater.721,Nat.Commun.8536,Nature.Nanotech.266, Sci.Rep.11055,Nat.Mater.179}.

A challenge that remains is to control magnetic anisotropy on a molecular level where the Heisenberg exchange and the single-ion anisotropy are the main governing mechanisms \cite{Nat.Chem.577,Nat.Phys.497,Science.988,J.Am.Chem.Soc.8694, Angew.Chem.Int.Ed.4413,Coord.Chem.Rev.1514,Angew.Chem.Int.Ed.2264,Inorg.Chem.2986, Chem.Eur.J.277,Phys.Rev.B.12177,Phys.Rev.Lett.066401,Inorg.Chem.10035}. The Heisenberg exchange interaction forces an alignment of neighboring spins, either parallel (ferromagnetic order) or antiparallel (antiferromagnetic order), 
creating a magnetic ground state without directional preference. However, when taking into account either spin-orbit, 
dipole-dipole, or antisymmetric exchange interaction 
a directional preference appears in the form of an anisotropic magnetization.


In order to study magnetic anisotropy from a fundamental point of view, we have synthesized a spin system containing four antiferromagnetically coupled Mn(II) ions \cite{Inorg.Chem.11903} and measured its magnetic properties using cantilever magnetometry. The availability of high magnetic fields (up to 33 T) enables us to access all the eleven possible spin states of the molecules and to fully map their anisotropic free-energy surface.  In particular, we will show that we can switch the magnetic easy and hard axis upon passing through the different states, which can be explained by a sign change in the single-ion anisotropy term of the spin Hamiltonian.

\begin{figure}[b] 
\includegraphics[width=0.8\columnwidth]{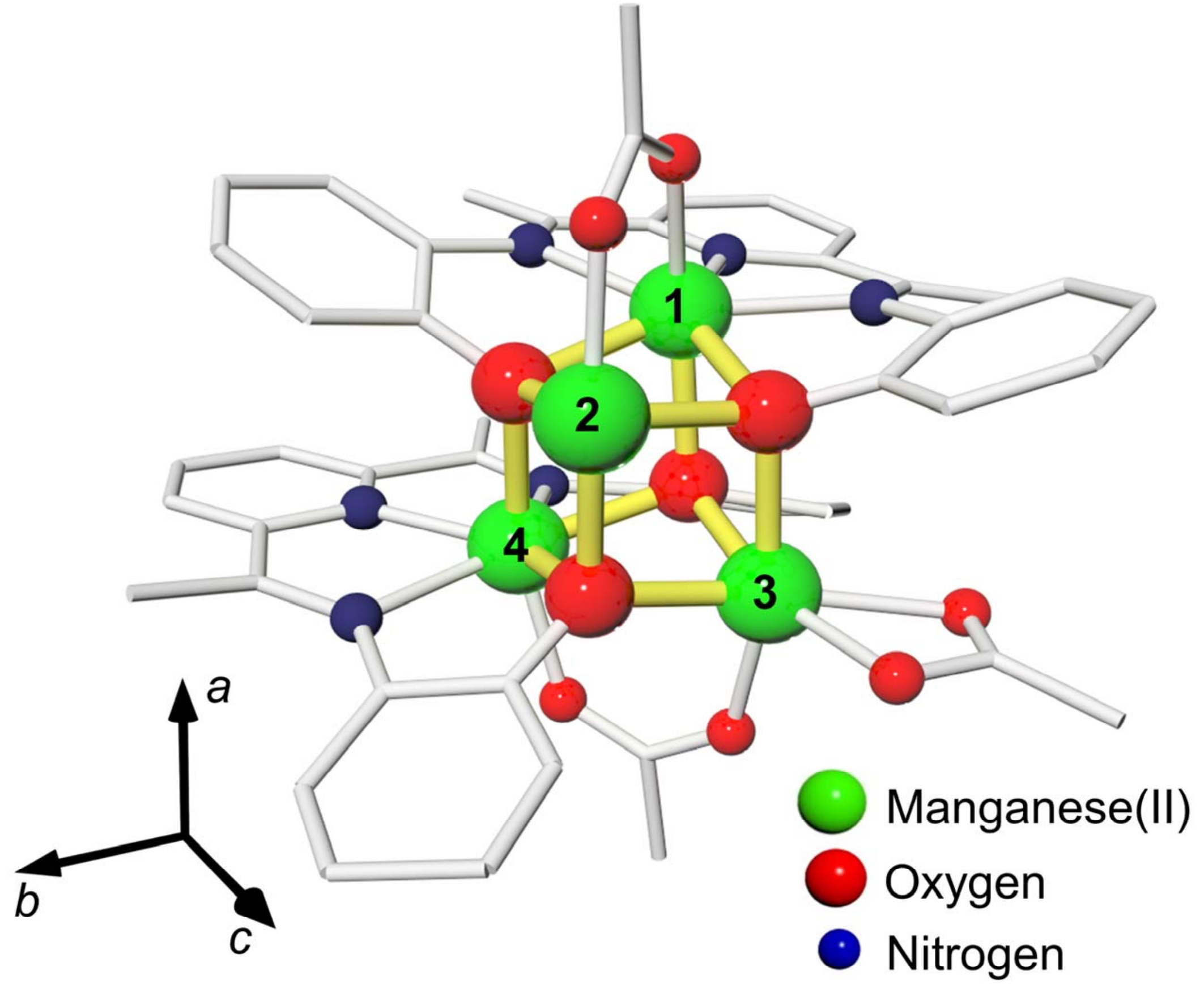}
\caption{\label{exp}
Structure of the Mn$_4$O$_4$ cluster as determined by single-crystal X-ray diffraction.
The carbon skeleton is shown as gray lines; a front-facing acetate on \textbf{2} 
and all hydrogen atoms are omitted for clarity. 
The four Mn(II) ions engage in an exchange interaction (yellow lines), which is mediated by orbital overlap between the $3d$ orbitals of the Mn(II) ions and the $2p$ orbitals of the diimine-pyridine ligand's oxygen atoms. 
}
\end{figure}

The Mn(II)$_4$O$_4$ cluster used for our experiments is shown in Fig.\,\ref{exp}. Millimeter-size single crystals were obtained using two-solvent diffusion, and characterized by single crystal X-ray diffraction. The crystal consists of clusters containing four antiferromagnetically coupled Mn(II) ions with equivalent ions \textbf{1},\textbf{4} and \textbf{2},\textbf{3}. More details on the synthesis and the structure determination can be found Ref.~\onlinecite{LaurensSchiff}.

The antiferromagnetic coupling between the four Mn ions is fine-tuned through the carboxylate ligands \cite{Inorg.Chem.11903}, resulting in a well-defined separation of all spin energies with $S=0$ as the magnetic ground state at zero magnetic field. The energetic separation to the $S=1, \ldots ,10$ higher-lying states is such that the system can be fully spin-polarized into its $S=10, M_S = -10$ state at the maximum available continuous magnetic field (33 T).



The magnetic properties of the crystal in a magnetic field are related to interactions at the atomic level through the effective 4-spin Hamiltonian \cite{Phys.Rev.B.214427}:
\begin{eqnarray}
\label{Hamiltonian}
\hat{\cal H} = -g\mu_B \vec{B}\cdot\hat{\vec{S}}
+ 2 \sum\limits_{i=1}^3 \sum\limits_{j=i+1}^4 J_{ij} \hat{\vec{s}}_i\cdot\hat{\vec{s}}_j
+\sum\limits_{i=1}^4 \hat{\vec{s}}_i\cdot\vec{\vec{D}}_i \hat{\vec{s}}_i 
\nonumber\\
+\frac{\mu_0 \; g^2 \mu_B^2}{4\pi}\sum\limits_{i=1}^3 \sum\limits_{j=i+1}^4
\left\{
\frac{3}{r^5_{ij}}[\vec{r}_{ij}\cdot\hat{\vec{s}}_i]
[\vec{r}_{ij}\cdot\hat{\vec{s}}_j] 
-\frac{1}{r^3_{ij}}
\hat{\vec{s}}_i\cdot\hat{\vec{s}}_j
\right\}.~~~
\end{eqnarray}
The first term is the Zeeman interaction between the magnetic field $\vec{B}$ and the total electron spin $\hat{\vec{S}}$ (i.e., the sum of spins $\hat{\vec{s}}_i$ of the four Mn ions), $\mu_B$ is the Bohr magneton and $g=2.0$ is the Land\'e $g$-factor of the Mn ions as determined by high-field EPR, see below.
The second term is the Heisenberg exchange interaction between the ions with antiferromagnetic coupling constants $J_{ij}<0$ determined from temperature-dependent magnetization measurements \cite{Inorg.Chem.11903}: $J_{13}=J_{14}=-2.2$ K,
$J_{12}=J_{13}=-1.1$ K and $J_{14}=J_{23}=-0.1$ K. 
The third term describes the single-ion anisotropy. The $3\times 3$  traceless and symmetric matrices $\vec{\vec{D}}_i$ are second rank tensors and depend on
five parameters for each ion. Since the cluster contains only two unique ions
({\bf 1}={\bf 4} and {\bf 2}={\bf 3}), the tensors $\vec{\vec{D}}_i$
of ions {\bf 4} and {\bf 3} are related by symmetry to those of
{\bf 1} and {\bf 2}, respectively. The last term contains the
magnetic dipole-dipole couplings between the ions. It depends
on the vectors $\vec{r}_{ij}$ that connect ions
$i$ and $j$ and the distances $r_{ij}=|\vec{r}_{ij}|$ and
has no free parameters;  $\mu_0$  is the vacuum permeability.

\begin{figure}[b]
\includegraphics[width=0.95\columnwidth]{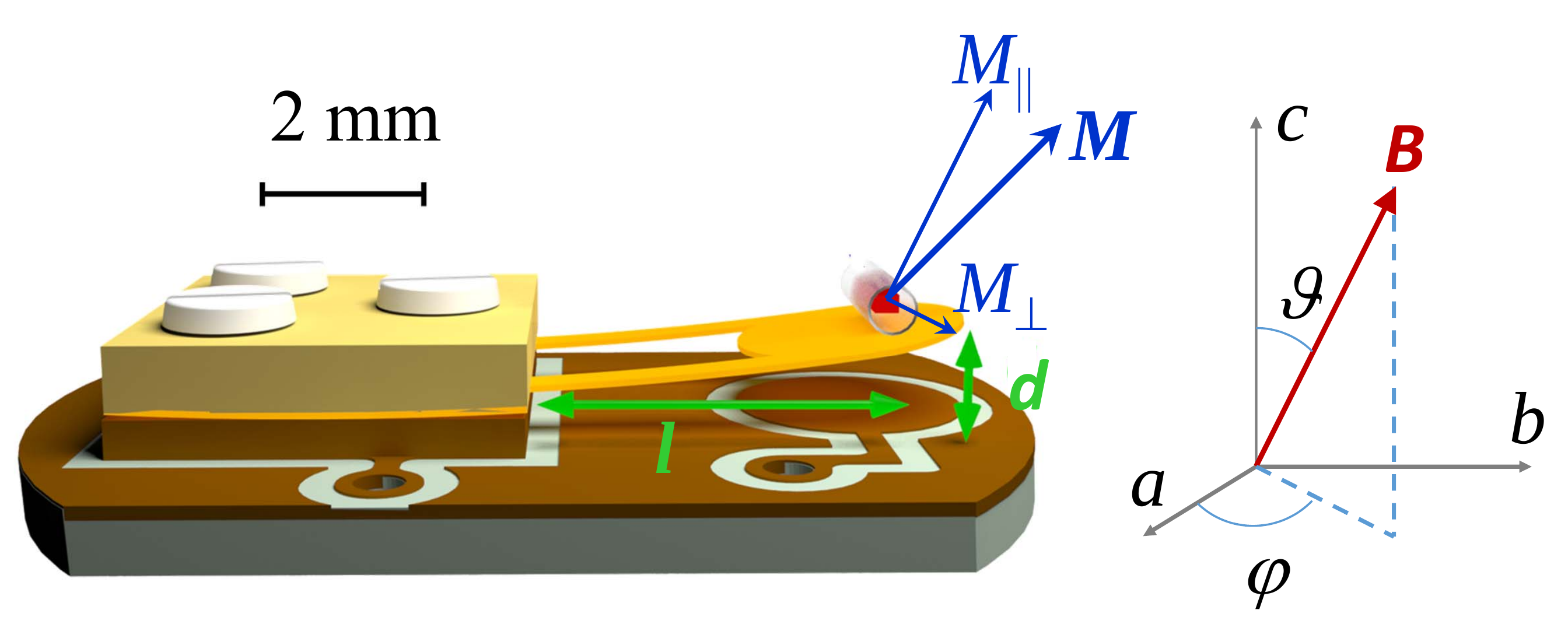}
\caption{\label{cantilever}
Representation of the cantilever setup used to measure the two vector components $M_\perp$ and $M_\parallel$  of the magnetization $\bf{M}$  in an applied magnetic field $\bf{B}$.
The right panel shows the orientation of the magnetic field and the definition of the angles $\vartheta$ and $\varphi$ with respect to the crystal axis.}
\end{figure}

When all parameters are known, Eq.\,(\ref{Hamiltonian}) can be used to calculate the free energy $F(B,\vartheta,\varphi)$ analytically from the eigenvalues $\epsilon_k$ of the Hamiltonian weighted by the temperature-dependent Boltzmann factors $\exp(-\epsilon_k/k_B T)$. For an anisotropic system, $F$ will depend on the absolute value of the magnetic field as well as its direction with respect to the crystal axes expressed by the angle $\vartheta$ between the $c$-axis and the field direction, and the polar angle $\varphi$ between the $a$ axis and the projection of the field in the $ab$-plane, see Fig.\,\ref{cantilever}.
The magnetization (in spherical coordinates) is then given by the vector derivative of the free energy with respect to the magnetic field:
\begin{equation}
\label{M}
\vec{M} = -\frac{\partial F}{\partial B} \hat{B} - \frac{1}{B} \frac{\partial F}{\partial \vartheta}  \hat{\vartheta}- \frac{1}{B \sin \vartheta} \frac{\partial F}{\partial \varphi} \hat{\varphi}
\end{equation}
with $\hat{B}$, $\hat{\vartheta}$ and $\hat{\varphi}$ the three spherical unit vectors corresponding to $\vec{B}$ .


\begin{figure}[t] 
\includegraphics[width=0.8\columnwidth]{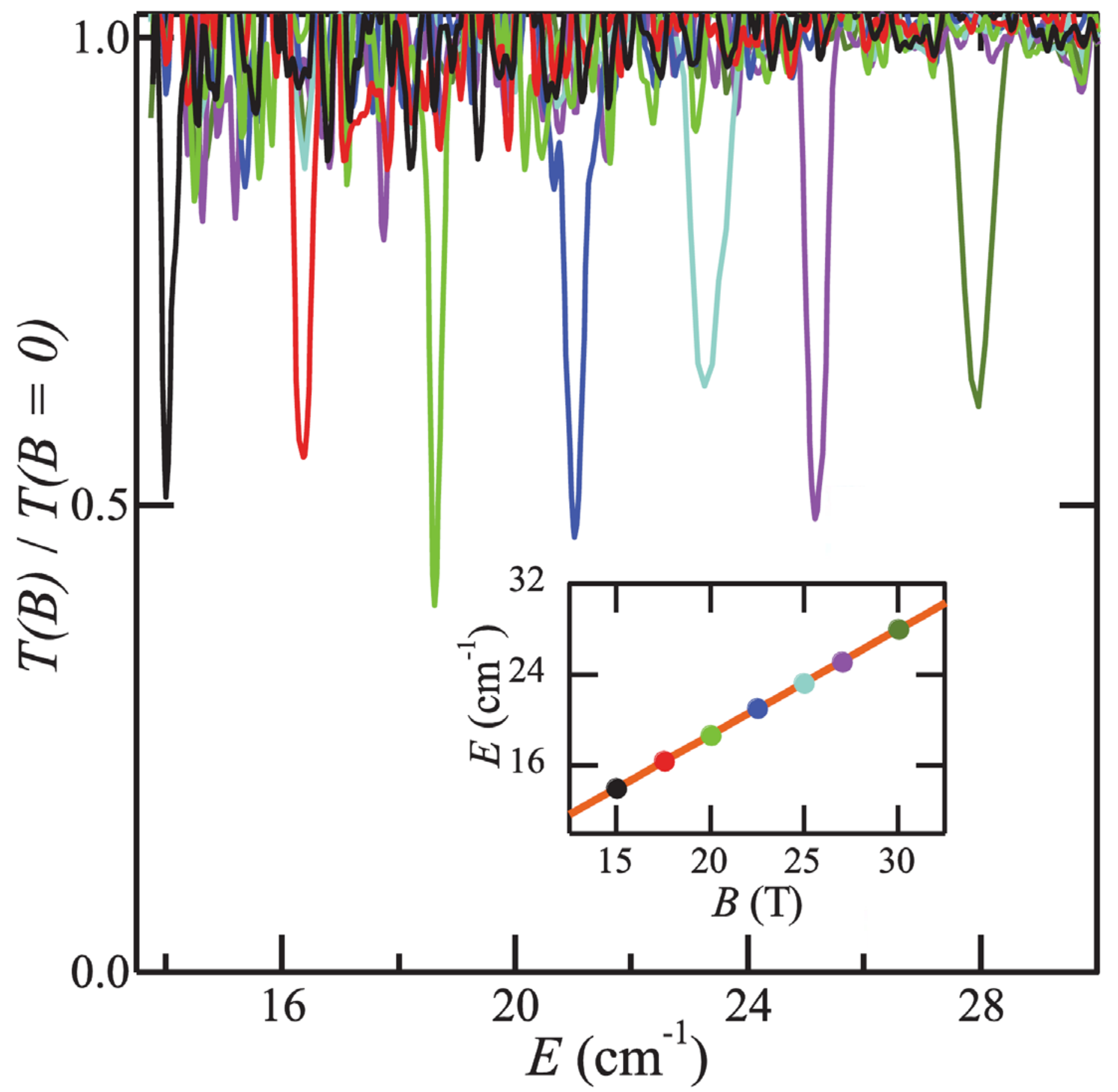}
\caption{\label{EPR}
Far-infrared  absorption of the Mn$_4$ clusters at 1.6 K at several constant magnetic fields
between 15 T (dark blue) and 30 T (dark green) \cite{ThesisErik}.
The inset shows the position of the absorption peaks as a function of magnetic field with 
the line  representing a linear Zeeman splitting corresponding to $g = 2.00$.
}
\end{figure}

In order to further characterize the $g$-factor in the Mn$_4$ clusters used in Eq.~\ref{Hamiltonian}, we have first performed high-field / high-frequency electron spin resonance measurements using a Bruker IFS113v
far-infrared Fourier spectrometer combined with a 33 T Bitter magnet \cite{ThesisErik}.  The sample was pressed in a pellet made from ground crystallites containing the Mn$_4$  clusters mixed with paraffin. 
The far-infrared transmission of the sample at $T=1.6$\,K  measured using a bolometer is shown in Fig.\,\ref{EPR}.
The instrumental resolution was set to 1 cm$^{-1}$; spectra were collected for 15 minutes and normalized to the zero magnetic-field background. The spectra recorded at different magnetic fields up to 30 T reveal a single absorption line and show that all
Zeeman levels are characterized by a free-electron Landé g-factor g = 2.00.

The core of this paper will be dedicated to a direct experimental access to the magnetic anisotropy of a molecular system.
For this, we have measured the magnetization vector using a home-built high-field rotational cantilever magnetometer. 
A sketch of the setup is shown in Fig.\,\ref{cantilever}: It consists of a flexible copper beryllium cantilever, to which we glued a glass capillary containing paraffin oil and an oriented single-crystal sample. 
Orientation of the sample with respect to the field and the cantilever has been performed using X-ray diffraction  with markings on the glass capillary indicating the crystallographic axes.
Rotating the sample holder, the magnetic field $\vec{B}$ can be applied  in any direction with respect to the platform normal and thus a given crystallographic direction. 

The sample develops a magnetization vector $\vec{M}$ containing a parallel component $M_\parallel$ along the field, and, for magnetically anisotropic systems, an additional perpendicular component $M_\perp$. When placed in a field gradient, the sample magnetization then leads to a total torque on the cantilever depending on both components: 
\begin{equation}  
\vec{\tau} = \vec{M} \times \vec{B} + \vec{\nabla}(\vec{M} \cdot \vec{B}) \times \vec{l} 
\end{equation}
with a lever arm $l$ defined by the distance between the sample and the torque axis. 

The first term is an anisotropic torque caused by $M_\perp$ and the second term arises from the force on $M_\parallel$ in a field gradient. 
By careful positioning of the cantilever in the field center, where $\vec{\nabla}(\vec{M} \cdot \vec{B})=0$, the capacitance change only originates from $\vec{M} \times \vec{B}$ (the torque). In a finite gradient, $\vec{M}_\parallel$ can be extracted by subtraction of the torque signal.

The torque can be deduced experimentally from the capacitance change between the cantilever (with the sample mounted on it) and a fixed back electrode. The capacitance was measured using an Andeen Hagerling 2700A capacitance bridge (15 V excitation, 10 kHz frequency). The setup was calibrated using an external DC voltage $V$  applied between the cantilever and the back electrode which yields a known torque $\tau \propto V^2$.

\begin{figure}[h]
\includegraphics[width=0.95\columnwidth]{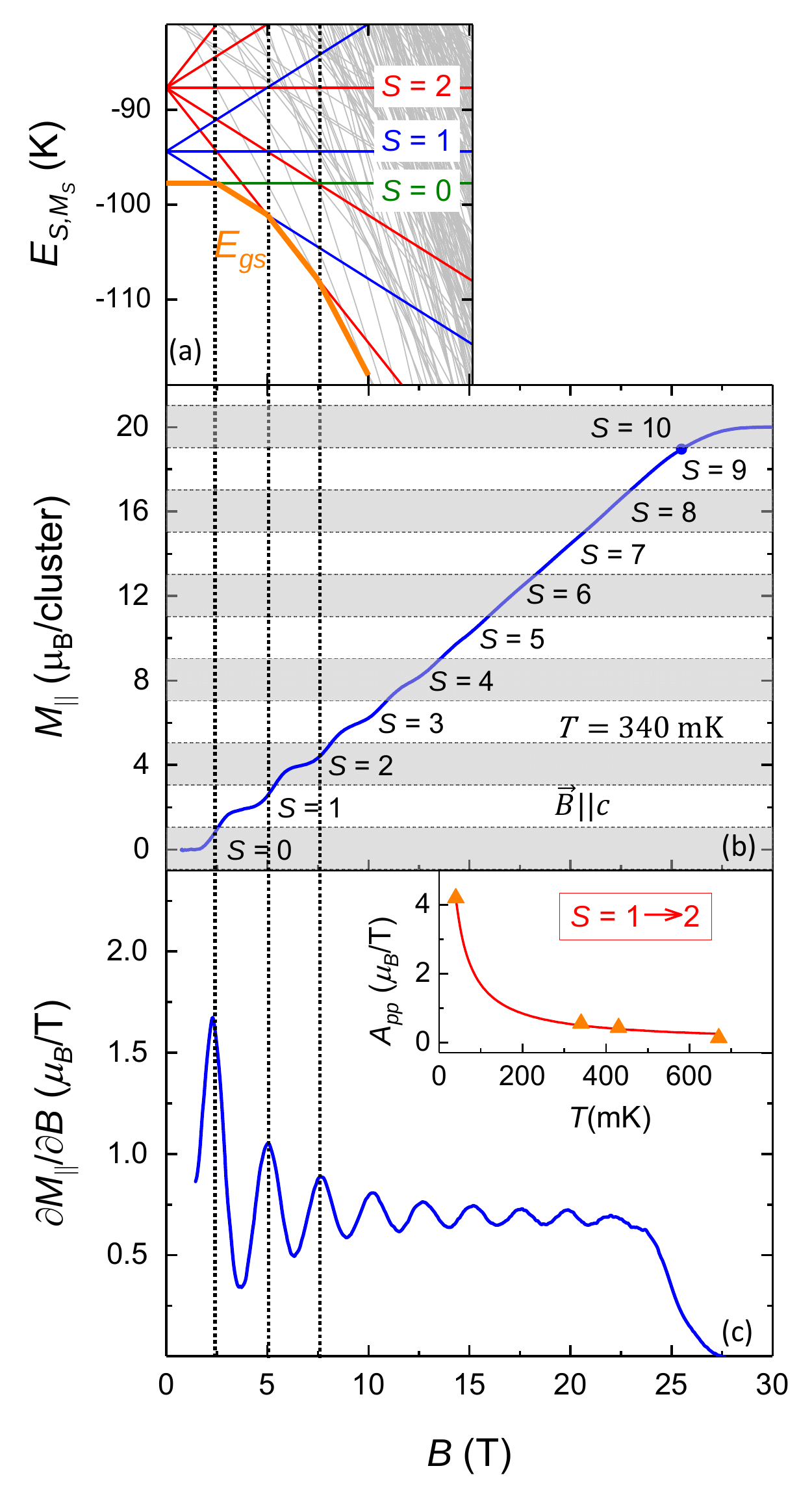}
\caption{\label{steps}
(a) Zeeman-split energy levels of the $S = 0, 1, \dots , 10$ states of four antiferromagnetically coupled Mn-ions as a function of the magnetic field. For clarity, we have omitted the dipole-dipole interaction in calculating this level scheme and we have not included  any (small) zero-field splitting effects of the individual spin states.  
Levels for $S= 0, 1, {\rm and~} 2$ are depicted in green, blue and red; higher spin states are shown in gray.
The ground state energy  is shown by the orange line.\newline 
(b) Step-wise increase of the parallel magnetization $M_\parallel$ of a single-crystal Mn$_4$O$_4$ cluster at $T=340$ mK.
The ground state changes from $S=0$ at low magnetic field to $S=10$ above 27 T.  \newline
(c) Field positions of the magnetization visualized as maxima in $\partial M_\parallel / \partial B$. 
The inset shows the  peak-to-peak amplitude of the second magnetization step characterized by an exponential decay as a function of temperature.
}
\end{figure}

The magnetization $M_\parallel$ of a single-crystal sample containing the Mn(II)$_4$O$_4$ clusters, measured at $T=340$ mK along the $c$-axis, is shown in Fig.\,\ref{steps}(b). 
It increases in a step-like fashion with fading steps at higher fields until it saturates at 27 T. The steps are less pronounced for higher magnetic fields, which is more clearly visible in its derivative $\partial M_\parallel / \partial B$ shown in Fig.\,\ref{steps}(c).  
The step positions, identified as maxima in $\partial M_\parallel / \partial B$, show an approximately regular $\sim 2.5$ T step interval. 
At higher temperatures, inset in Fig.\,\ref{steps}(c), the steps become smeared out with the amplitude of $\partial M_\parallel / \partial B$ following the expected exponential behavior \cite{Phys.Rev.B.174440}.

\clearpage

The steps in $M_\parallel$ can be attributed to abrupt changes of the field dependence of the ground state energy \cite{J.Appl.Phys.5090,J.Appl.Phys.4155,J.Phys.Soc.Japan.1178}. The four Mn(II) ions contain a total of twenty unpaired $d$-electrons, resulting in eleven possible spin-states $S=0, 1, \ldots , 10$. At zero field, these states are energetically separated mainly by the Heisenberg exchange interaction. Upon applying a magnetic field, each state splits into $2S+1$ Zeeman levels with $M_S=-S$ as the lowest energy state, see Fig.\,\ref{steps}(a).
The Zeeman effect lowers the energy of these levels by $g\mu_B M_S B$ and two states cross at approximately 2.5~T intervals. The first step appears at 2.5 T when ($S=1,\;M_S=-1$) crosses ($S=0,\;M_S=0$) and becomes the ground state; the final step at 25 T is associated with the crossing of ($S=10,\;M_S=-10$) with ($S=9,\;M_S=-9$). Generally, the  ($S,\;M_S$) levels consist of several individual coupled spin states with an additional fine structure in their energy splitting. For example, six states associated to (0,0) are crossing fifteen states associated to (1,-1) at  2.5 T and three states associated to (9,-9)  are crossing one (10,-10) state at 25 T. However, since we only observe ten individual magnetization steps, we are not able to resolve this fine structure experimentally and it is therefore reasonable to assume that their energy splitting is small compared to the splitting between different spin states depicted in Fig.\,\ref{steps}(a).
These level crossings entail  magnetization-steps of $2\mu_B$ in $M_\parallel$. 
At higher magnetic fields, due to the single-ion anisotropy and magnetic dipole-dipole interactions, the crossings of adjacent spin-states in Fig.\,\ref{steps}(a)  become avoided crossings, leading to less pronounced steps
 \cite{Phys.Rev.B.052407,J.Chem.Phys.3361,J.Appl.Phys.7822,Phys.Rev.B.06403,Phys.Rev.B.174440,Phys.Rev.Lett.096403}.

Observing such well-isolated states makes it possible to select any of the spin states as the ground state, and, as we will show in the following, to address their magnetic anisotropy individually.
We probe this anisotropy by placing the sample into the field center (with no field gradient) and measuring the torque $M \times \vec{B}$, which is then purely defined by the anisotropic magnetization component $M_\perp$  dependent on the relative orientation of the field.


\begin{figure}[b] 
\includegraphics[width=0.8\columnwidth]{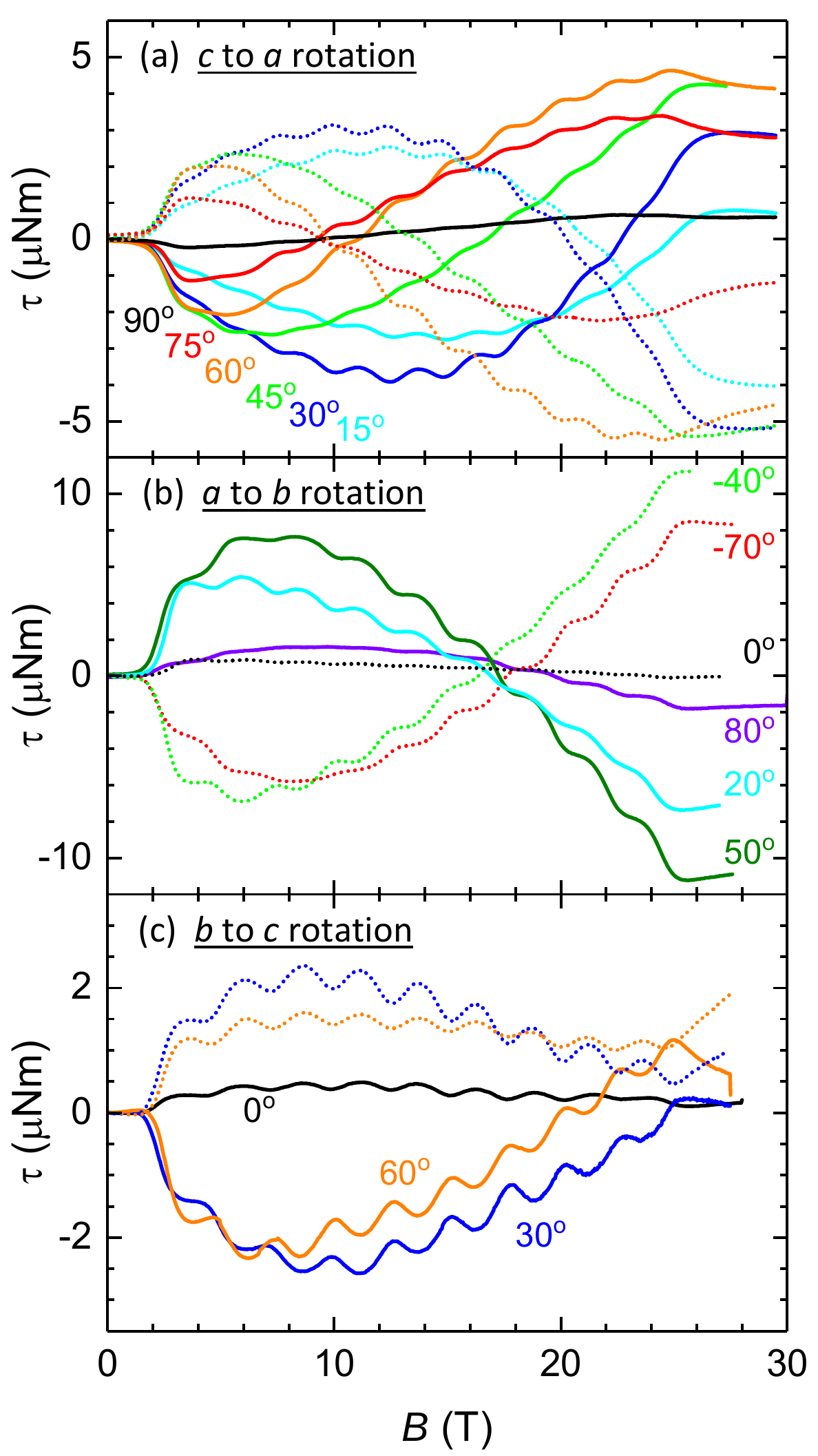}
\caption{\label{torque-and-steps}
Torque measured at  $T=340$ mK, while rotating the direction of the magnetic field from $c$ to $a$ (a), $a$ to $b$ (b) and $b$ to $c$ (c).
The colors represent the rotation angle, dotted lines show corresponding negative angles. 
Sign changes in the torque represent a change of the anisotropic easy and hard axis (see main text). }
\end{figure}

The directional dependence of the torque as a function of magnetic field when rotating the field from the $c$-axis towards the $a$-axis is shown in Fig.\ref{torque-and-steps}(a). Similar to the magnetization steps, we observe oscillations in the torque (i.e., in the perpendicular magnetization component $M_\perp$)  caused by the change in ground state energy. For low magnetic fields and positive rotation angles, the observed torque is negative, i.e.,~towards the $c$-axis, which defines this axis  as an easy  axis in the $ac$-plane. 
Interestingly,  the torque changes sign in higher magnetic fields between $10$ and $27$~T, strongly dependent on the angle between the crystallographic axis and the field.  This sign change reflects a change in the magnetic anisotropy upon increasing the field, i.e.,~passing through the individual spin states $S=0, 1, \ldots , 10$.

A similar behavior can also be observed in the rotation from $a$ to $b$  in Fig.\,\ref{torque-and-steps}(b), where the $a$-axis changes from a hard axis at low magnetic fields to an easy axis at high magnetic fields. When rotating from $b$ to $c$, see Fig.\,\ref{torque-and-steps}(c), the anisotropy is much smaller and the complex form of the free-energy surface in the $bc$-plane is reflected in the torque.

\begin{figure}[t]  
\includegraphics[width=0.95\columnwidth]{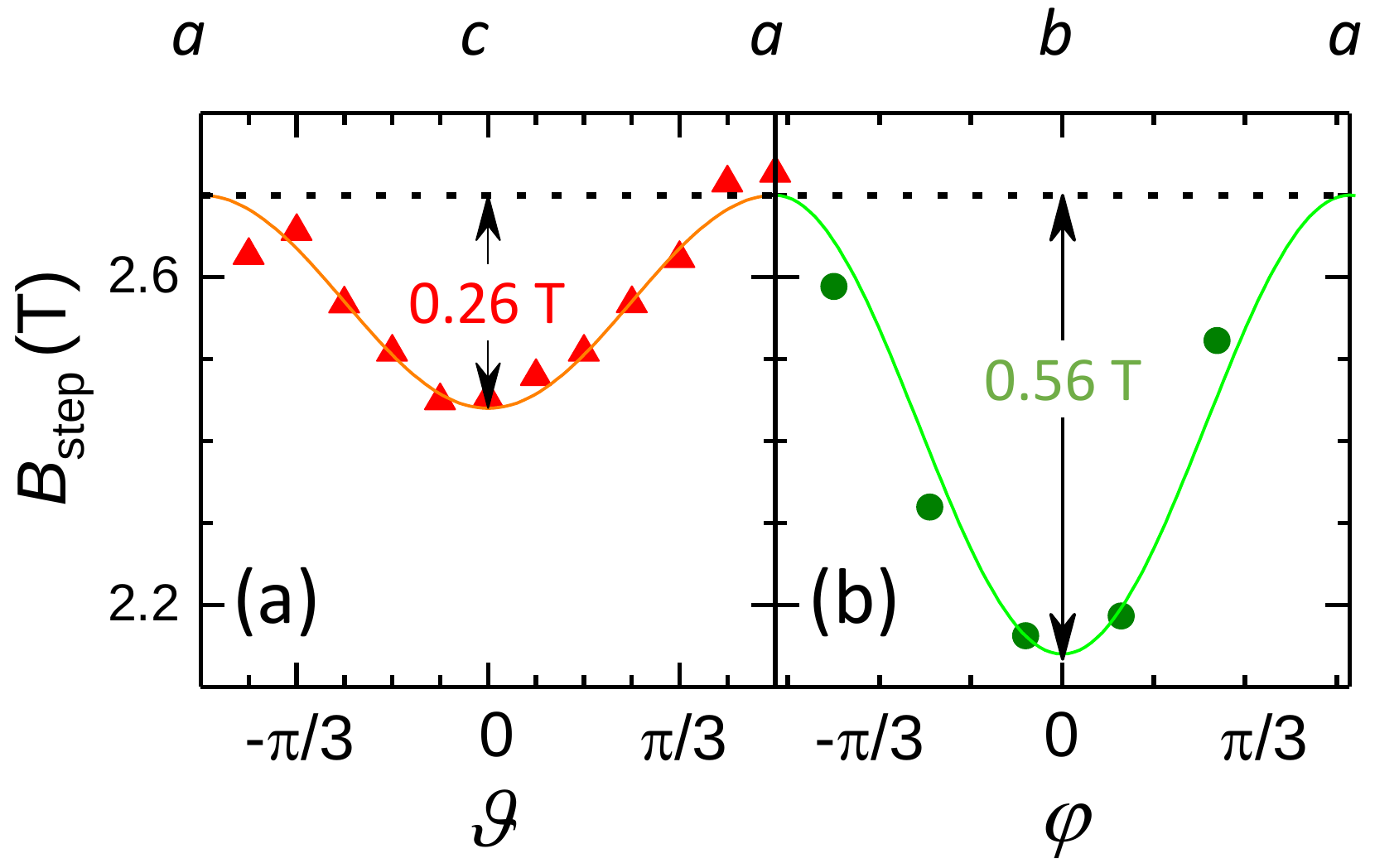}
\caption{\label{steppos}
Angular dependence of the field position of the $S=0 \rightarrow 1$ step for rotations from  $c$ to $a$ (b) and from $b$ to $a$ (c), respectively. 
}
\end{figure}

\begin{figure}[h] 
\includegraphics[width=0.99\columnwidth]{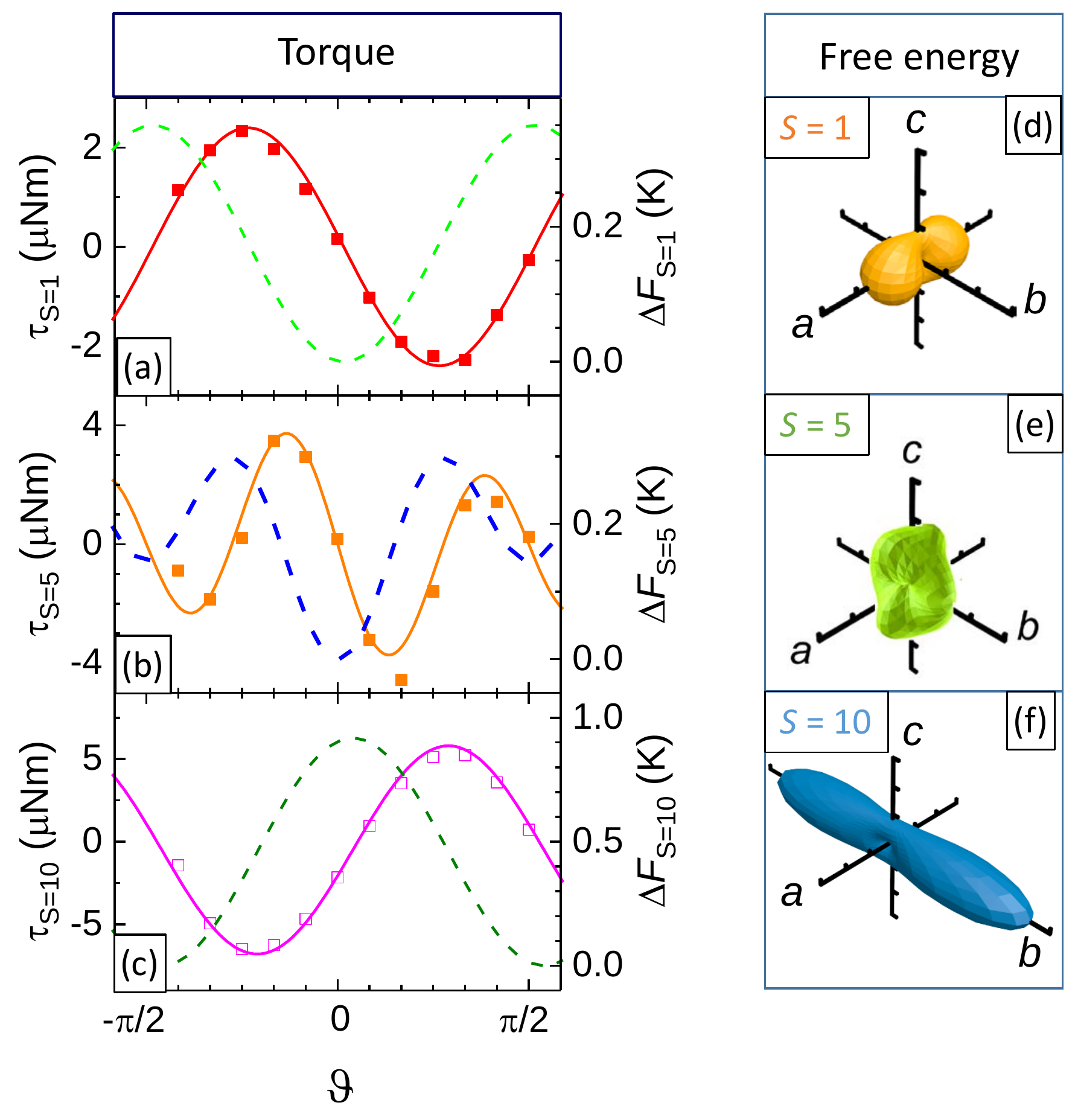}
\caption{\label{angle}
Observed angular dependence of the torque (lines and symbols, left y-axis) and the change in free energy (dashed lines, right y-axis)) for three selected spin states. 
The right panels depict their free energy surfaces expressed as a function of the direction of the applied magnetic field with respect to the crystallographic directions.}
\end{figure}

The change in anisotropy when passing from one spin state to the next one is also visible in the field positions of the magnetization steps. Indeed, the step positions directly reflect the angular dependence of the free-energy changes between two different spin states.
In particular, since the $S=0$ state is isotropic, the $S =0 \rightarrow 1$  step provides a quantitative access to the free-energy anisotropy of the $S=1$ state which occurs when the $M_S=-1$ level of $S=1$ crosses the $S=0$ state and becomes the ground state, see Fig.\,\ref{steps}(a).  
When rotating from $c$ to $a$, see Fig.\,\ref{steppos}(a), the step position increases by $\Delta B = 0.26$~T. 
With $g=2.0$ this yields an anisotropy barrier, i.e.,~a change of free energy, $\Delta F = 0.35$~K. 
Analogously, the change of 0.56 T on step-position when rotating from $b$ to $a$  (Fig.\,\ref{steppos}(b)) yields an anisotropy barrier of 0.75 K. 
Using these values it is possible to reconstruct the angular-dependent free energy surface of the $S=1$  state quantitatively from the experimental data;
the result is shown in Fig.\,\ref{angle}(d) \cite{note-multiple-states}.


Since for $S>1$ the step positions only reflect the relative changes in the free-energy surface between two spin states, it is more convenient to use the experimentally measured torque for a given state as a more direct access to its anisotropy. Specifically, measuring the torque for all principal rotations ($a$ to $c$, $a$ to $b$, and $b$ to $c$) and integrating it over the angle allows to reconstruct  $F(\vartheta ,\varphi)$. 
This is shown explicitly for the angular dependence of the torque at constant magnetic field for three given spin states $S=1$, $S=5$, and $S=10$ in the left panels of  Fig.\,\ref{angle}.
For $S=1$ ($B=3.7~$T) the free energy is minimum along the $c$-axis  leading to a negative torque for positive angles. In contrast, for $S=10$ ($B=26~$T)  the sign of the torque is reversed, the $c$-axis is now a hard axis and the free energy is minimum along the $a$-axis. Analogously, rotating from $b$ to $a$ (not shown) allows us to identify a
free-energy minimum (easy axis) along $b$ for $S=1$ which becomes a maximum (hard axis) for $S=10$. Using the absolute calibration of the $S=1$ free-energy surface performed above, we can also calculate the absolute values for the magnetization anisotropy of the $S=10$ state and plot the corresponding free energy surface, see Fig.\,\ref{angle}(f).  The easy axis is now along $a$ and the anisotropy barriers along $b$ and $c$ are now 2.93 K  and 0.74 K, respectively.

Cycling through the spin-states, the anisotropy changes and we are able to follow the complete evolution of the anisotropic free energy surface for each individual spin state.
In particular, the anisotropy switches axis around $S=5$, yielding a complex free-energy surface as illustrated in Fig.\,\ref{angle}(e).  These observations show that magnetic anisotropy is not only determined by structure, but also by the individual spin state which is populated and can therefore be controlled without changing the structure.





Finally, it is worth mentioning that the qualitative features of the magnetic field dependent anisotropy as represented by the free energy surfaces in Fig.\,\ref{angle}(d-f) follow a remarkably simple rule.  The dominant contribution to the anisotropy arises from the axial anisotropy in the single ion terms as expressed by the parameters $D_i=D_{zz,i}-[D_{xx,i}+D_{yy,i}]/2$, where the $x$, $y$, and $z$ subscripts refer to the components of the second rank tensors $\vec{\vec{D}}_i$ in the principal axes frames of ion $i$. This contribution changes sign when passing through the $S =0, 1, \ldots 10$ spin states leading to an anisotropy reversal.  Although hints of this general trend have been observed before \cite{Phys.Rev.Lett.066401,Phys.Rev.Lett.096403}, the ability to access and analyze the magnetic anisotropy of each individual spin states, makes this now abundantly clearer.

In conclusion, we have shown that we can control and switch magnetic anisotropy on a molecular level by the magnetic-field driven occupation of distinct spin states. An observed anisotropy reversal can be tracked down to a single-ion term which changes sign when passing from the lowest to the highest spin state. 

\vspace*{5em}

\newpage

\begin{acknowledgments}
This work has been supported by HFML-RU/NWO-I, member of the European Magnetic Field Laboratory (EMFL).
It is part of the research programme of the Stichting voor Fundamenteel Onderzoek der Materie (FOM, which is financially supported by the Nederlandse Organisatie voor Wetenschappelijk Onderzoek (NWO).
We thank Jan van Leusen (RWTH Aachen) for enlightening theoretical discussions.
\end{acknowledgments}


\begin{thebibliography}{32}
\bibitem [{not()}]{note-multiple-states}%
  \BibitemOpen
  \href@noop {} {}\bibinfo {note} {As mentioned above, the
  individual ($S, M_S$) levels consist of several spin combinations from the
  four Mn atoms, though such an additional fine structure is not resolved
  experimentally in additional magnetization steps. Therefore, here and in the
  following we only access the anisotropy of all states associated to a
  specified ($S, M_S$).}\BibitemShut {Stop}%
\end{thebibliography}


%










\end{document}